\documentclass[12pt]{article}

\usepackage[dvips]{graphicx}
\usepackage{epsfig}
\usepackage{amsmath,amsfonts,amssymb,amsthm}
\usepackage{verbatim}
\usepackage{psfrag}
\usepackage{bm}
\usepackage{bbm}
\usepackage[square,comma,sort&compress,numbers]{natbib}
\usepackage{color}
\usepackage{slashed}

\usepackage{epsf,epsfig}
\usepackage{graphics}

\newcommand{\lsim}
{\;\raisebox{-.3em}{$\stackrel{\displaystyle <}{\sim}$}\;}

\begin{document}
\thispagestyle{empty}

\begin{flushright}
{
\small
TUM-HEP-938-14\\ CERN-PH-TH/2014-059
}
\end{flushright}

\vspace{0.4cm}
\begin{center}
\Large\bf\boldmath
Spectator Effects during Leptogenesis in the Strong Washout Regime
\unboldmath
\end{center}

\vspace{0.4cm}

\begin{center}
{Bj\"orn~Garbrecht$^a$ and Pedro Schwaller$^b$}\\
\vskip0.2cm
{\it $^a$Physik Department T70, James-Franck-Stra{\ss}e,\\
Technische Universit\"at M\"unchen, 85748 Garching, Germany\\
\vspace{0.15cm}
$^b$CERN, Theory Division, CH-1211 Geneva 23, Switzerland}\\
\vskip1.4cm
\end{center}

\begin{abstract}
By including spectator fields into the Boltzmann equations for Leptogenesis, we show that partially equilibrated spectator interactions can have a significant impact on the freeze-out value of the asymmetry in the strong washout regime. The final asymmetry is typically increased, since partially equilibrated spectators "hide" a part of the asymmetry from washout. We study examples with leptonic and non-leptonic spectator processes, assuming thermal initial conditions, and find up to 50\% enhanced asymmetries compared to the limit of fully equilibrated spectators.  Together with a comprehensive overview of the equilibration temperatures for various Standard Model processes, the numerical results indicate the ranges when the limiting cases of either
fully equilibrated or negligible spectator fields are applicable and when they are not. Our findings also indicate an increased sensitivity to initial conditions and finite density corrections even in the strong washout regime. 
%
%
\end{abstract}



\section{Introduction}

In standard Leptogenesis, the asymmetry is created
from out-of-equilibrium decays of heavy singlet neutrinos into lepton
and Higgs doublets~\cite{Fukugita:1986hr}.
The Standard Model (SM) encompasses a whole cascade of interactions that
equilibrate in the Early Universe subsequently
at temperatures below $\sim 10^{15}\, {\rm GeV}$. Many of these
processes can redistribute the initial asymmetries to
spectator degrees of freedom, see {\it e.g.} Refs.~\cite{Nardi:2005hs,Nardi:2006fx} for an overview.
The spectator particles do not directly couple to the singlet neutrinos and
are therefore not directly washed out. This can have a sizeable impact
on the freeze-out value of the lepton asymmetry~\cite{Barbieri:1999ma,Buchmuller:2001sr,Nardi:2005hs,Nardi:2006fx,Abada:2006fw}.

\begin{figure}[t]
\begin{center}
\includegraphics[width=0.7\textwidth]{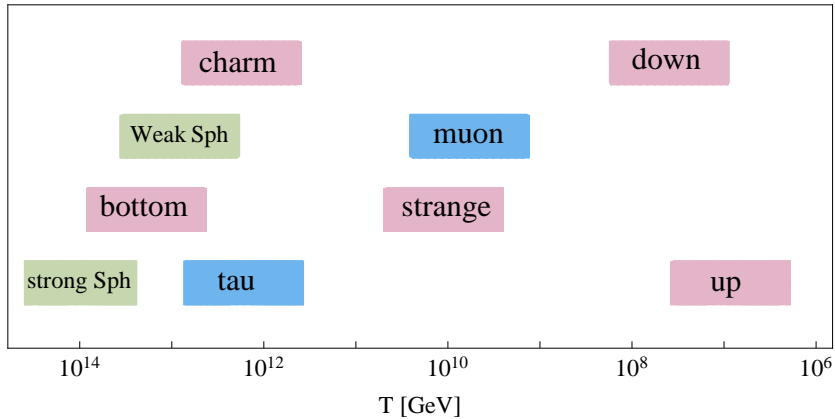}
\caption{Temperature ranges where different SM interactions equilibrate. Shown are the strong and weak sphalerons (green), and quark (red) and lepton (blue) Yukawa interactions. The bands range from $T_X$ to $20\,T_X$, with $T_X$ denoting the equilibration temperatures for the individual interactions that are calculated in more detail in Section~\ref{section:regimes}. 
The bands indicate the range where one can expect nontrivial effects due to partial equilibration, based on the results of Sections~\ref{sec:partial:tauequilibration} and \ref{sec:Example:Leptogenesis}.}
\label{fig:regions}
\end{center}
\end{figure}

As more spectator fields reach chemical equilibrium, the washout rate is more
reduced, leading to an increase in the efficiency of the production of the 
asymmetry~\cite{Nardi:2005hs,Nardi:2006fx}. The rates of the numerous SM interactions
that mediate chemical equilibrium are spread over a large temperature range, 
which we illustrate in Figure~\ref{fig:regions}. One should therefore
expect that situations when spectator fields partially equilibrate (as
opposed to the limiting cases when they are either negligible or fully equilibrated) are not exceptional.
Quantifying the limitations of the approximations of negligible or fully equilibrated
spectator effects and the proper treatment of the intermediate regime is the main purpose
of the present work.

When shifting the freeze-out temperature to lower values,
one may expect that the efficiency of Leptogenesis simply increases monotonously, because
the spectator fields move more closely to chemical equilibrium. While for some regions of parameter
space, in particular when washout is not too strong, this is indeed the case,
there is however an interesting curiosity (that can be observed for the partial equilibration of $\tau$-Yukawa interactions in the numerical analysis of Ref.~\cite{Beneke:2010dz}) that
generically occurs for stronger washout:
The freeze-out asymmetry for partially equilibrated spectator fields can be enhanced
compared to the two limiting cases. This is because in the strong washout regime, large
asymmetries are present at early times, that are then partially transferred to the
spectator sector, where these are ``hidden'' from washout.
At later times, in particular when washout processes freeze out, these asymmetries can exceed
the asymmetries present in the case of full equilibration (where the spectators closely
track the asymmetry in left-handed leptons, that experiences direct washout).
In this work, the effect is quantified and explained in some detail on numerical examples
for Leptogenesis. Interestingly, the intermediate enhancement can be present in either case,
whether the partially equilibrated
spectator processes are leptonic or non-leptonic.

The outline of this paper is as follows: In Section~\ref{sec:setup},
we review the well-known Boltzmann equations for Leptogenesis
in the presence of spectator effects, mainly for the purpose of
introducing our notations and conventions. The
various temperature scales where the equilibration of particular
SM processes occur are discussed in Section~\ref{section:regimes}.
As for the Yukawa-mediated processes, we make use of perturbative
techniques and results that
were recently presented in Refs.~\cite{Anisimov:2010gy,Besak:2012qm,Garbrecht:2013bia}, while the rates of the instanton-mediated
processes are taken or deduced from lattice simulations reported in Refs.~\cite{Bodeker:1998hm,Moore:2000mx}. The results derived in Refs.~\cite{Anisimov:2010gy,Besak:2012qm,Garbrecht:2013bia} for leptonic
processes are generalised to interactions involving quarks in
Appendix~\ref{appendix:b:rate}.
As an example of the effect that is the main topic of this work, we calculate the
impact of partial equilibration of $\tau$-lepton Yukawa couplings in
Section~\ref{sec:partial:tauequilibration}. This is of interest, as one
should expect here the largest effect from the partial equilibration on the
freeze-out asymmetry, as part of the asymmetry (in principle up to one third)
is hidden from washout
in the right-handed $\tau$-leptons. However, in the present work, we neglect effects from
partial flavour decoherence, that should be included in future work along the
lines of Ref.~\cite{Beneke:2010dz}. Instead in Section~\ref{sec:Example:Leptogenesis} we consider a high temperature example, the equilibration of strong and weak sphalerons and $b$-quark Yukawa interactions, in a temperature regime where lepton flavour effects are negligible. It turns out that even in the absence of flavour effects, parts of the asymmetry can be hidden from washout in the right-handed $b$-quarks and in lepton doublets that are not directly affected by washout. 
Finally, we summarise and discuss the results in
Section~\ref{sec:conclusions}.

\section{Boltzmann Equations for Leptogenesis}
\label{sec:setup}
In this Section, we review the standard Boltzmann equations
for Leptogenesis that are
valid when interactions with spectator fields are either negligible (interaction
rates much smaller than the Hubble rate) or  when these are fully equilibrated
(rates much above the Hubble rate). One purpose of this discussion is
to introduce notations and conventions used in the present paper.
The equations presented here are generalised
to account for partial equilibration of the spectator processes in the subsequent Sections.

The Boltzmann equations usually are derived by substituting cross sections from Quantum
Field Theory (QFT) into the collision term. An important caveat in that approach is
that a subtraction of real intermediate states is necessary in order to
reproduce a unitary evolution of the system, such that no asymmetry is generated
in equilibrium, as it is required by the $CPT$ (charge-parity-time) theorem.
Alternatively, the Boltzmann equations for Leptogenesis can be derived from the
Schwinger-Dyson equations of Non-Equilibrium QFT~\cite{Garny:2009rv,Beneke:2010wd,Anisimov:2010dk}, which are unitary
by construction. The resulting equations from both methods agree in the
limit of the strong washout regime,
what is our main concern here. The methods of obtaining solutions for
the freeze-out lepton asymmetries and the properties
of these are discussed in detail in Ref.~\cite{Buchmuller:2004nz}.

In order to account for the expansion of the Universe, it is useful to introduce
the conformal time $\eta$, that is related to the comoving time $t$ as
$dt=a\,d\eta$, where the scale factor in a radiation-dominated Universe is
$a=a_{\rm R}\eta$. We take
\begin{align}
a_{\rm R} = T_{\rm com} = \frac{m_{\rm Pl}}{2} \sqrt{\frac{45}{g_\star \pi^3}} \,,
\end{align}
where $m_{\rm Pl} = 1.22 \times 10^{19}~{\rm GeV}$ is the Planck mass, $g_\star = 106.75$ is the number of relativistic degrees of freedom and
$T_{\rm com}$ is the comoving temperature (in terms of which the physical
temperature is $T=T_{\rm com}/a$). By this choice, the comoving time is simply given by
$\eta=1/T$. Now, an asymmetry in left-handed SM leptons $\ell$ is generated from
the decay of sterile right-handed neutrinos $N_i$. We take the masses of $M_i$ to
be hierarchical, {\it i.e.} $M_1\ll M_2, M_3$. When considering the
out-of-equilibrium dynamics of $N_1$ only,
the time evolution can be conveniently
parametrised through the dimensionless variable $z = M_1/T$, such that
the Hubble expansion rate is given by $H(z) = a_{\rm R}^{-1} M_1^2/z^2$. 

We denote asymmetries in a particle species $X$ as $q_X = n_X - n_{\bar X}$, where the charge densities $q_X$ and number densities $n_X$ refer to a single gauge degree of freedom. For small deviations from chemical equilibrium, the charges are related to the chemical potentials via $q_X = T^2/6 \mu_X$ for a Weyl fermion and $q_X = T^2/3 \mu_X$ for a complex scalar. Finally it is useful to introduce entropy normalised quantities $Y_X = q_X/s$, where $s = g_\star (2 \pi^2/45) T^3$. Since the sterile neutrinos are Majorana particles,
these must be described in terms of their number densities $n_{Ni}$
instead of charge densities. We define $n_{Ni}$ to include the internal degrees of
freedom, {\it i.e.} as the sum over the two helicity components, and use the
entropy normalised number densities $Y_{Ni} = n_{Ni}/s$.

When spectator and flavour effects are neglected, the Boltzmann equations for
Leptogenesis are given by
\begin{subequations}
\label{BEQs:paradigm}
\begin{align}
	\frac{d}{dz}  Y_{N1}  & = - \bar{{\cal C}}_{N1} \left(Y_{N1} - Y^{\rm eq}_{N1} \right), \label{eqn:eom1}\\
	\frac{d}{dz} Y_\ell & = \frac12\bar{S} \left(Y_{N1} - Y^{\rm eq}_{N1} \right) - \bar{W} \left(Y_\ell + \frac{1}{2} Y_H \right),
	\label{eqn:eom2}
\end{align}
\end{subequations}
where $\bar {\cal C}$ is the thermally averaged decay rate of $N_1$, $\bar S$  the
source term and $\bar W$ the washout term.

For the simplest scenarios of Leptogenesis and
in the strong washout regime, the freeze-out asymmetry is independent of the
evolution at early times, and thus approximations that are valid for $z>1$
can be applied: One may treat $N_1$ in the non-relativistic limit and employ
Maxwell distributions instead of Fermi-Dirac and Bose-Einstein distributions.
This results in the following expressions for the various thermally averaged rates:
\begin{subequations}
\label{av:rates}
\begin{align}
	\bar{\cal C}_{N1} & = |Y_1|^2 \frac{z T_{\rm com}}{8 \pi M_1} \,, \\
	\label{S:average}
	\bar{S} & = \varepsilon \,\bar{\cal C}_{N1} \qquad \text{with}\quad \varepsilon = \frac{\Gamma_{N_1 \to \ell H}-\Gamma_{N_1 \to \bar\ell H^\dagger}}{\Gamma_{N_1 \to \ell H}+\Gamma_{N_1 \to \bar\ell H^\dagger}} \,, \\ 
	\bar{W} & = |Y_1|^2 \frac{3 T_{\rm com} e^{-z} z^{\frac{5}{2}}}{2^{\frac{7}{2}} \pi^{\frac{5}{2}} M_1} \,, \\
	Y_{N1}^{\rm eq} & = 2\,\frac{T_{\rm com}^3z^2}{2 \pi^2} K_2(z)\frac{1}{s} \approx {2^{-\frac 1 2} \pi^{-\frac 3 2}}z^{\frac{3}{2}} e^{-z}\frac{T_{\rm com}^3}{s} \,,
\end{align}
\end{subequations}
where $Y_1$ is the Yukawa coupling of $N_1$ to $\ell$ and $H$, and $K_2(z)$ is a Bessel function. Note that in general,
$\ell$ is a linear combination of the three SM lepton doublets.
In Eq.~(\ref{S:average}), we employ the usual definition of the decay asymmetry
$\varepsilon$, where $\Gamma_{N_1 \to \ell H}$ ($\Gamma_{N_1 \to \bar\ell H^\dagger}$)
denotes the partial decay rate of $N_1$ into leptons (antileptons).
Note that $\bar{{\cal C}}_{N1}$ is the thermal average of the total decay
rate of $N_1$, whereas $Y_\ell$ accounts for one component of the gauge multiplet only.
This is the origin of the explicit factor $\frac12$ in front of the source term
in Eq.~(\ref{eqn:eom2}).

A key quantity controlling the dynamics of Leptogenesis is
\begin{align}
\label{K:1}
K_1=\bar{\cal C}_{N1}(z=1)\,.
\end{align}
The basic regimes for Leptogenesis are typically separated
between strong and weak washout, defined by $K_1>1$ and $K_1<1$, respectively.
In the strong washout regime, to a good approximation, the lepton asymmetry freezes out
when $z\approx z_{\rm f}$, where~\cite{Buchmuller:2004nz}
\begin{align}
\label{z:f}
z_{\rm f}\approx 1 + 0.5 \log\left(1 + 0.0031 K_1^2 \log^5\left(9.6 K_1^2\right)\right)\,.
\end{align}
To guide the eye in our presentation of the numerical results in Figures~\ref{fig:evolution} and~\ref{fig:strongsphbYukawa} below, we quote the values
$z_{\rm f}(K_1=\{1,10,10^2,10^3,10^4\})=\{1.1,5.2,8.8,12.0,14.9\}$
and also refer to the graph of that function that is presented in Ref.~\cite{Buchmuller:2004nz}.

Since in the strong washout regime and for thermal initial
conditions of the sterile neutrinos, $(Y_{N1} - Y_{N1}^{\rm eq})\ll Y_{N1}^{\rm eq}$,
the solution to Eq.~(\ref{eqn:eom2}) is to a good approximation
\begin{align}
\label{DeltaYN}
Y_{N1} - Y_{N1}^{\rm eq} & = \frac{1}{\bar{\cal C}_{N1}(z) }\frac{d}{dz} Y_{N1}^{\rm eq}(z) \,,
\end{align}
which we employ for the numerical results presented in this paper. This relation also
holds for non-thermal initial conditions provided $z\gg 1$. As we see below, the outcome
of Leptogenesis with partially equilibrated spectators to some extent also depends on
contributions from $z\lsim 1$, which is why we restrict the analysis in this paper to
the case of thermal initial conditions. Note that inaccuracies in the calculation
of $\bar{\cal C}_{N1}$ cancel when substituting the deviation from equilibrium~(\ref{DeltaYN}) into the source term~(\ref{S:average}). Remaining uncertainties
concern the decay asymmetry $\varepsilon$, but as this quantity is seen to increase
towards small values of $z$ in Refs.~\cite{Garny:2009rv,Beneke:2010wd}, our approximations for the enhancement of the
freeze-out asymmetry from partially equilibrated spectators should be conservative.

Now, the
interactions that can change $Y_\ell$ and $Y_H$ without changing the baryon-minus-lepton asymmetry $Y_{B-L}$ are usually referred to as spectator effects. 
In order to take account of these, Eq.~(\ref{eqn:eom2})
must be accordingly modified. The condition of a vanishing weak hypercharge
in conjunction with equilibrium conditions can be used to determine the
value of $Y_H$.
Furthermore, lepton number is not conserved when the weak sphalerons come into thermal equilibrium, and one should instead use $\Delta=B-L$ as a conserved charge, which is only broken by interactions mediated by the $N_i$. 

The usual treatment of spectator effects is to assume that a certain interaction is either fully equilibrated or not at all. Then it is straightforward to write down an evolution equation equation for $Y_{B-L}\equiv Y_\Delta=\Delta/s$ in the form~\cite{Nardi:2005hs}
\begin{align}
\label{evo:lepto}
	\frac{d}{dz} Y_\Delta & = -\bar{S} \left(Y_{N1} - Y^{\rm eq}_{N1} \right) + 2 \bar{W} \left(c_\ell + \frac{1}{2} c_H \right) Y_\Delta\,,
\end{align}
where the coefficients $c_\ell$, $c_H$ are defined via $Y_\ell = c_\ell Y_\Delta$, $Y_H = c_H Y_\Delta$ and can be determined using the equilibrium conditions (see Section~\ref{sec:Example:Leptogenesis} for more details). Note that $Y_\Delta$ is defined including the sum over gauge degrees of freedom. When at lower temperatures
interactions mediated by lepton Yukawa couplings are in equilibrium, one
must distinguish between the particular asymmetries
$\Delta_i=B/3-L_i$, where $i=e,\mu,\tau$~\cite{Barbieri:1999ma,Nardi:2006fx,Abada:2006fw}.
When arranging the $Y_{\Delta i}$ in a column vector, $c_\ell$ and $c_H$ can be written
as a matrix and a row-vector in lepton flavour space, respectively. This is
the usual approach pursued in calculations on flavoured Leptogenesis~\cite{Nardi:2006fx,Abada:2006fw}.
In case of incomplete flavour equilibration, one should take account of the
correlations between the different lepton flavours, as it is described
in Refs.~\cite{Beneke:2010dz,Blanchet:2011xq}.
Combining the partial flavour decoherence with partial
equilibration of spectator effects will be the topic of future work.
In the examples that we discuss in this paper, we neglect the flavour effects.

The approach of assuming either complete equilibration of spectators or no equilibration
at all immediately leads to the question when precisely these limits are valid and it
requires the knowledge of the rates for the spectator processes. An overview
of these rates and references to their derivations is presented in the following Section.

\section{Rates for Spectator Processes}
\label{section:regimes}

In the following, we review the interactions that equilibrate at time scales relevant for Leptogenesis, i.e. for temperatures $T\lesssim 10^{14}\,{\rm GeV}$, and that can mediate
spectator processes. In particular these are the strong and weak sphalerons and the Yukawa interactions of various quarks and leptons. Gauge and top Yukawa interactions equilibrate above these scales and will not be considered here. 

The method of obtaining the sphaleron rate at high temperatures
is developed in Ref.~\cite{Bodeker:1998hm}. It is applied to the SM,
in particular including the effects of a Higgs boson, in Ref.~\cite{Moore:2000mx},
where it is found that the diffusion rate for the Chern-Simons number is
\begin{align}
\label{Eq:sphrate1}
\Gamma_{\rm sph}=(8.24\pm0.10)\left(\frac{N}{2}\right)^5
\frac{g^2T^2}{m_{\rm D}^2}\left({\log\frac{m_{\rm D}}{g^2T}}+3.041\right)\alpha^5 T^4\,,
\end{align}
where $g$ is the ${\rm SU}(N)$ gauge coupling,
$\alpha=g^2/(4\pi)$ and $m_{\rm D}$ the Debye mass of the
${\rm SU}(N)$ gauge bosons.
The numerical coefficients have strictly been derived for $N=2$, while
the explicit scaling $\sim N^5$ is conjectured in
Ref.~\cite{Moore:2000ara}, {\it cf.} the comments below.
In the SM at high temperatures, the Debye mass-square of the  $W$ bosons is
$m_{\rm D}^2=\frac{11}{6}g_2^2 T^2$, and we then obtain
the weak sphaleron rate $\Gamma_{\rm ws}=\Gamma_{\rm sph}|_{g=g_2}$, where
we set $g_2=0.55$ as suggested by renormalisation group running for temperatures
$\sim 10^{12}\;{\rm GeV}$.
There is also recent work that reports the sphaleron rate in the SM during the
crossover between the high-temperature regime and low temperatures,
where the electroweak symmetry is broken~\cite{D'Onofrio:2012jk}.
The Chern-Simons number diffusion rate enters the kinetic equations in the
form~\cite{D'Onofrio:2012jk,Khlebnikov:1988sr,Burnier:2005hp}
\begin{align}
\label{sph:kin}
\frac{d (B/3+L_i)}{d t}=-\frac{1}{T} \Gamma_{\rm ws}
\sum\limits_{i=1}^{N_{\rm f}}\left(3\mu_{Qii}+\mu_{Lii}\right)\,,
\end{align}
where $N_{\rm f}=3$ is the number of flavours in the SM.
See also~\cite{Bento:2003jv} for a simple and pragmatic discussion on how to substitute $\Gamma_{\rm ws}$ into the kinetic equations.
On the left hand side, we may as well replace the charge densities $B$ and $L_i$ 
with $6q_{Qii}$ and $2q_{Lii}$. As stated in Section~\ref{sec:setup} above,
we use the convention that $q_X$ denotes
the charge density of a species $X$ of a particular entry of a gauge multiplet.
Furthermore, we allow in this Section the charge densities and chemical potentials to
be matrices, as this allows to account for possible correlations in flavour space.
This is done in view of future work where we aim to include decohering
flavour correlations. In the remainder of this paper, we set the flavour correlations
to be zero and denote the diagonal components of the flavour correlation matrices
using a single index, {\it i.e.} $q_{Xi}=q_{Xii}.$

Using these definitions and relations, we can break
Eq.~(\ref{sph:kin}) down to
\begin{subequations}
\begin{align}
\frac{d q_{Qij}}{d t}=&-\frac{\Gamma_{\rm ws}}{2 T^3} \delta_{ij}
\left(9 {\rm tr}[q_{Q}]+ 3{\rm tr}[q_{L}]\right)\,,
\\
\frac{d q_{Lij}}{d t}=&-\frac{\Gamma_{\rm ws}}{2 T^3} \delta_{ij}
\left(9 {\rm tr}[q_{Q}]+3{\rm tr}[q_{L}]\right)\,,
\end{align}
\end{subequations}
where the traces are taken over flavour space. The non-zero eigenvalues
associated with this homogeneous system of differential equations
are $18\Gamma_{\rm ws}/T^3$.
Taking $g_2\approx0.55$ for the ${\rm SU}(2)$ gauge coupling of the
SM around $10^{12}\,{\rm GeV}$, we find that $18\Gamma_{\rm ws}/T^3= H$
for { $T= T_{\rm ws}\approx 1.8\times 10^{12}{\rm GeV}$}. In the absence of other interactions this would be the weak sphaleron relaxation temperature, but including top Yukawa and strong sphalerons as spectators reduces the relevant temperature by a factor of two~\cite{Bento:2003jv}. 

For the strong ${\rm SU}(3)$ interactions, the sphaleron rate is perhaps less well known 
than for the weak ${\rm SU}(2)$ case.
The question is addressed in Refs.\cite{Moore:1997im,Moore:2000ara}, where
the more recent of these suggests that the sphaleron rate in ${\rm SU}(N)$
theory scales as $\sim N^5$. The Debye mass-square is now given by
$m_{\rm D}^2=2g_3^2 T^2$ and we take $g_3=0.61$ for $T\sim 10^{12}\,{\rm GeV}$. 
Now, the strong sphalerons
contribute to the kinetic equations as
\begin{align}
\label{eqn:strongSph}
\frac{d q_{Qij}}{dt}=-\frac{d q_{u ij}}{dt}=-\frac{d q_{d ij}}{dt}
=-\frac{1}{3}\frac{\Gamma_{\rm ss}}{ T^3}\delta_{ij}
\left(6{\rm tr}[q_Q]-3{\rm tr}[q_u]-3{\rm tr}[q_d]\right),
\end{align}
where $\Gamma_{\rm ss}=\Gamma_{\rm sph}|_{g=g_3}$. 
Here the non-zero eigenvalue associated with the system of equations is $12\Gamma_{\rm ss}/T^3$. Equilibration should therefore occur when $12\Gamma_{\rm ss}/T^3=H$, which implies an equilibration temperature of $T_{\rm ss}\approx 2.4\times 10^{13}\,{\rm GeV}$. Again this is slightly reduced when the top Yukawa is taken into account.

The calculation of the Yukawa-mediated rates at high temperatures
requires a proper account of $2\leftrightarrow 2$ processes
involving gauge radiation. In the hot plasma, these rates turn out to be dominated
by the $t$-channel exchange of fermions.
For the production of light singlet fermions,
a calculation is reported in Refs.~\cite{Anisimov:2010gy,Besak:2012qm}.
Using CTP techniques, the results are confirmed in Ref.~\cite{Garbrecht:2013bia}, where
in addition, also  the rates involving the leptonic SM Yukawa couplings are
obtained. It is found that
\begin{subequations}
\label{eqn:flavour_t}
\begin{align}
\frac{d q_{L}}{dt}=-\frac{\gamma^{{\rm fl}\delta \ell}}{2}(h^\dagger h q_L + q_L h^\dagger h)+\gamma^{{\rm fl}\delta \ell}h^\dagger q_R h\,,
\\
\frac{d q_{R}}{dt}=-\gamma^{{\rm fl}\delta \ell}(h h^\dagger q_R + q_R h h^\dagger)+2 \gamma^{{\rm fl}\delta \ell}h q_L h^\dagger\,,
\end{align}
\end{subequations}
where
$\gamma^{{\rm fl}\delta \ell}\approx 5\times 10^{-3}T$ (see also Ref.~\cite
{Cline:1993bd}
for an earlier estimate).
The Yukawa matrix $h$ is diagonal in the mass eigenbasis of the charged leptons.
The equilibration temperature for the $\tau$ flavour,
where $h_{\tau\tau}^2\gamma^{{\rm fl}\delta \ell}=H$ is then given by
$T_\tau\approx 3.7\times 10^{11}{\rm GeV}$. The corresponding temperature
for $\mu$ leptons is given through $h_{\mu\mu}^2\gamma^{{\rm fl}\delta \ell}=H$
as $T_\mu\approx 1.3\times 10^{9}\,{\rm GeV}$. 

In turn, the rate for equilibration of the up and down quark Yukawa interactions
has not yet been studied in detail. As it is
explained in Appendix~\ref{appendix:b:rate}, the intermediate results
presented in Ref.~\cite{Garbrecht:2013bia} can however be employed to
obtain this rate as $\gamma^{{\rm fl}\delta d}\approx \gamma^{{\rm fl}\delta u}\approx 1.0\times 10^{-2}T$
(the superscripts $d/u$ indicate the down/up-type quark), at temperatures $T\sim 10^{12}~{\rm GeV}$. For $b$-quarks, this
implies an equilibration temperature of $T_b=4.2\times 10^{12}\,{\rm GeV}$. While the running of couplings can be mostly neglected for the charged lepton Yukawa interactions, the effect is notable as $\gamma^{{\rm fl} \delta d}$ increases to $1.2\times 10^{-2}T$ at $T\sim 10^9~{\rm GeV}$ (strange quark Yukawa equilibration) and to $1.5\times 10^{-2}T$ at $T\sim 10^6~{\rm GeV}$ (down/up equilibration). 

\begin{table}
\begin{center}
\begin{tabular}{|r|l|l|}
\hline
$T_{\rm ss}$ & $2.4\times10^{13}\,{\rm GeV}$ & strong sphaleron\\
$T_{\rm ws}$ & $1.8\times10^{12}\,{\rm GeV}$ & weak sphaleron\\
$T_{b}$ & $4.2\times10^{12}\,{\rm GeV}$ & bottom-quark Yukawa\\
$T_{c}$ & $3.8\times10^{11}\,{\rm GeV}$ & charm-quark Yukawa\\
$T_{s}$ & $2.5\times10^9\,{\rm GeV}$ & strange-quark Yukawa\\
$T_{u}$ & $1.9\times10^6\,{\rm GeV}$ & up-quark Yukawa\\
$T_{d}$ & $8.8\times10^6\,{\rm GeV}$ & down-quark Yukawa\\
$T_{\tau}$ & $3.7\times 10^{11}\,{\rm GeV}$ & $\tau$-lepton Yukawa\\
$T_{\mu}$ & $1.3\times 10^9\,{\rm GeV}$ & $\mu$-lepton Yukawa\\
$T_{e}$ & $3.1\times 10^4\,{\rm GeV}$ & electron Yukawa\\
\hline
\end{tabular}
\end{center}
\caption{\label{tab:rates}Equilibration temperatures $T_X$ for Yukawa- and instanton-mediated SM processes.
Methods of the calculations and uncertainties are discussed in the text. Partial equilibration is relevant when the freeze-out of the lepton asymmetry happens at temperatures between $T_X$ and $20\;T_X$. }
\end{table}

A summary of the various equilibration temperatures $T_X$,
where the interaction rates agree with
the Hubble rate, is given in Table~\ref{tab:rates}. It should be noted that the equilibration temperatures are affected by spectators (in a similar way as the washout), so the temperatures given in the table should only serve as guidelines. 
A pictorial overview
is presented in Figure~\ref{fig:regions}, where we show the bands ranging
from $T_X$ to $20\; T_X$. The choice of the location and widths of the bands
depends of course on the requirements on the precision of the Boltzmann equations. 
By comparing with the numerical examples in the subsequent Sections, one should be
able to see this in more detail, and one may conclude that partial equilibration
of spectator effects is not an exceptional situation.

In the literature (see e.g.~\cite{Nardi:2005hs,Nardi:2006fx}), it is often assumed that the strong sphalerons equilibrate before bottom and tau Yukawas, and that the weak sphalerons equilibrate before second generation Yukawas. Here instead we note that it seems more appropriate to assume that the weak sphaleron equilibrates after the bottom but before tau Yukawas, and that the strong sphaleron might not be fully equilibrated when the bottom Yukawa becomes relevant. Overall it seems that in most regions more than one spectator interaction can be partially equilibrated. Noting these partial discrepancies,
one should however keep in mind systematic uncertainties, in particular
in the determination of the strong sphaleron rate.

\section{Partial $\tau$ Yukawa Equilibration During Leptogenesis}
\label{sec:partial:tauequilibration}
\begin{figure}[t]
\begin{center}
\includegraphics[width=0.48\textwidth]{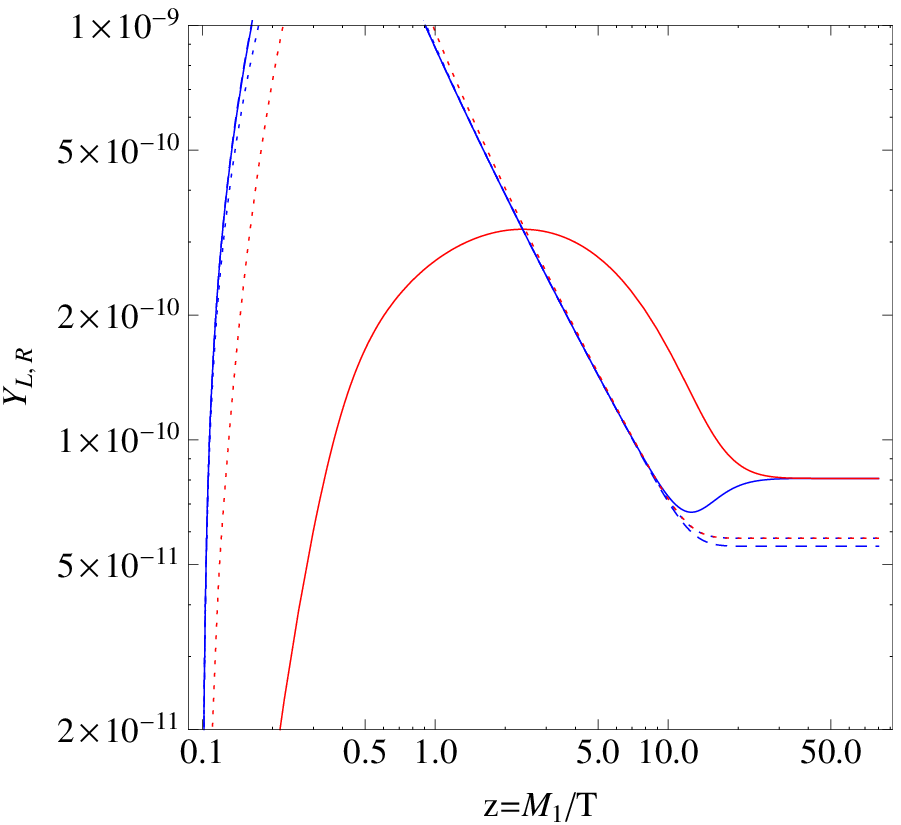}
\hspace*{.5cm}
\includegraphics[width=0.44\textwidth]{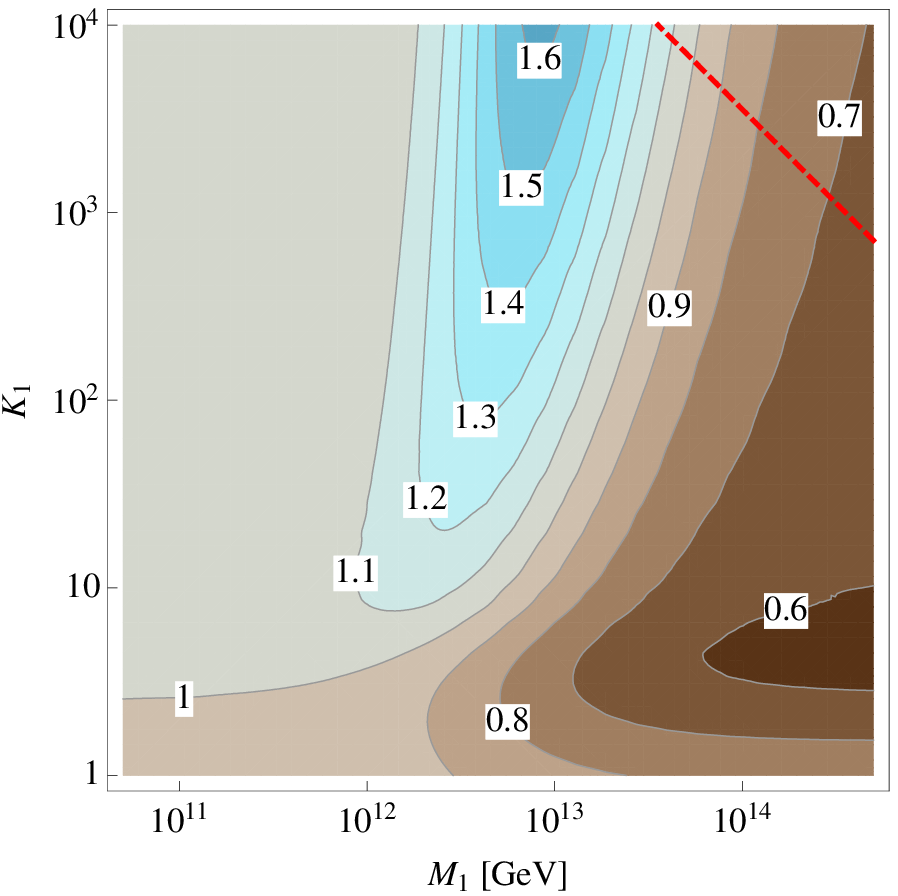}
\caption{Left: Solution of the evolution equations~(\ref{eqn:flavour1}) for $M_1=4\times 10^{12}~{\rm GeV}$ and $Y_1 = 0.3$. Shown are solutions for $h_\tau = \sqrt2 m_\tau/v$ (solid), for $h_\tau =0$ (dashed) and for fully equilibrated tau Yukawa, $h_\tau\to\infty$, (dotted). Right: Enhancement of the asymmetry from partial spectator equilibration. 
Shown is the ratio of the freeze-out asymmetry $2Y_\ell+Y_{\rm R}$ for the
case $h_\tau = \sqrt2 m_\tau/v$ divided by the case $h_\tau\to\infty$, 
as function of the lightest right-handed neutrino mass $M_1$ and of the washout strength $K_1$ as defined in Eq.~(\ref{K:1}). As expected, in the low and high $M_1$ limits this ratio goes to unity and $2/3$, respectively. 
Finally the region above the dashed red line would require non-perturbatively large Yukawa couplings ($Y_1^2 \geq 4 \pi$) to realise the strong washout, and is therefore unphysical. 
}
\label{fig:evolution}
\end{center}
\end{figure}

For Leptogenesis in the strong washout regime, a comparably
large lepton asymmetry is present in the plasma around $z\sim1$, after which it is suppressed from washout until the freezes out occurs at $z\sim 10-20$, {\it cf.} the dashed line in Figure~\ref{fig:evolution} for example. Spectator effects can transfer parts of the asymmetry into particles that are not subject to washout, like right-handed charged leptons. If the spectator interaction is fully equilibrated, the asymmetry in spectator fields closely tracks the lepton doublet asymmetry, such that this effect is adequately described by a modified washout factor, as implemented in Eq.~(\ref{evo:lepto}). 

However this description is not valid if the spectator effects equilibrate during the same times the lepton asymmetry freezes out. In that case, parts of the large asymmetry around $z\sim 1$ can be transferred to the spectator fields, but they will only fully equilibrate with the lepton doublet after the washout ends, thus \textit{hiding} parts of the asymmetry from washout. As we show in the following, this effect can have a significant numerical impact on the predicted final asymmetry, compared to both the cases where spectator effects are neglected or assumed to be fully equilibrated.

For illustration, we consider here first a simplified
scenario where the $\tau$-Yukawa coupling equilibrates during Leptogenesis. A part of the lepton asymmetry is directly
transferred to right-handed $\tau$-leptons and hidden from washout processes.
We expect the resulting effect to be larger than for non-leptonic spectator
processes, where part of the hypercharge asymmetry that is initially present in Higgs bosons is distributed over a larger number of quark degrees of freedom.
For simplicity, we ignore here other fully equilibrated spectator processes as well as flavour effects and the partial equilibration
of weak sphalerons, that should occur at the same time. Therefore, the results do
not serve the purpose of a full phenomenological study, but will rather indicate the
size of the effect and for which range of parameters it may be important. A more realistic
scenario is presented in the next Section.

To account for the partial equilibration of left handed and right-handed
leptons $\ell$ and ${\rm R}$,
the Boltzmann equations for the leptons~(\ref{eqn:eom2}) or~(\ref{evo:lepto}) are now replaced by
\begin{subequations}\label{eqn:flavour1}
\begin{align}
\frac{d Y_{\ell}}{dz}&=\frac{\bar{S}}{2} \left(Y_{N1} - Y^{\rm eq}_{N1} \right) - \bar{W} Y_\ell -h_\tau^2\gamma^{{\rm fl}\delta \ell} \frac{T_{\rm com}}{T M_1}(Y_\ell-Y_{\rm R})\,,
\\
\frac{d Y_{\rm R}}{dz}&=-2 h_\tau^2 \gamma^{{\rm fl}\delta \ell}\frac{T_{\rm com}}{T M_1}(Y_{\rm R}-Y_\ell)\,,
\end{align}
\end{subequations}
where $Y_{\ell,{\rm R}} = q_{\ell,{\rm R}}/s$ denote the asymmetries in a single gauge component, as before,
and where $h_\tau\equiv h_{\tau\tau}$. Note that compared to Eqs.~(\ref{eqn:flavour_t}) a factor $T_{\rm com}/(T M_1)$ arises from the change of the time variable from $dt$ to $dz$. This also cancels the explicit $T$ dependence in $\gamma^{{\rm fl}\delta \ell}$. 
The evolution of $N_1$ is still determined by Eq.~(\ref{eqn:eom1}) and the approximate solution~(\ref{DeltaYN}).

Numerical solutions to Eqs.~(\ref{eqn:flavour1}) are presented in Figure~\ref{fig:evolution}. In the left panel, we show the time evolution of the asymmetries
$Y_{\ell,{\rm R}}$
for one particular parameter point with the SM value of the lepton-Yukawa coupling
\begin{align}
h_\tau=\sqrt2 m_\tau/v\,,
\end{align}
where $m_\tau$ is the mass of the $\tau$-lepton $v=246\,{\rm GeV}$
and
$M_1 = 4\times 10^{12}~{\rm GeV}$ and $Y_1 = 0.3$ (solid lines). For comparison, there are also the evolutions of the asymmetry in the absence of tau Yukawa interactions, $h_\tau=0$, (dashed lines) and the evolutions in the fully
equilibrated case, $h_\tau\to\infty$ (dotted lines). In all cases, the freeze out of the
asymmetries is clearly visible around $z=10$.
For $h_\tau\not=0$, we see that the right-handed asymmetry (red) is not equilibrated fast enough to follow the asymmetry in the lepton doublets (blue). It slowly increases until $Y_\ell$ drops below $Y_{\rm R}$, and  it is much larger than $Y_\ell$ at the time where the washout processes freeze out. The left-handed asymmetry is then replenished until $Y_\ell = Y_{\rm R}$, and is about 40\% larger than in the case where the tau Yukawa is assumed to be fully equilibrated (dotted lines), and where the 
right-handed asymmetry closely follows the one in the left-handed doublet. 

In the right panel of Figure~\ref{fig:evolution}, we show the magnitude of this effect as a function of the Leptogenesis scale and of the washout strength $K_1$ as given in Eq.~(\ref{K:1}). The asymmetry $2Y_\ell+Y_{\rm R}$ obtained from the solution to Eq.~(\ref{eqn:flavour1}) for $h_\tau=\sqrt2 m_\tau/v$
is divided by the asymmetry in the limit where the $\tau$ Yukawa interactions are fully equilibrated,
$h_\tau\to\infty$. As expected, this ratio goes to unity for $M_1 \ll T_\tau$, while it goes to $2/3$ for $M_1 \gg T_\tau$ where the asymmetry is only distributed among the two left-handed degrees of freedom. More importantly, we note that the asymmetry can be enhanced for intermediate values of $M_1$ compared to these limiting cases. While the enhancement is strongest in the regime where the washout is very large, it is still sizeable in regions with moderately strong washout of $K_1 \sim 10-100$. Furthermore it should be noted that the region where neither approximation gives an accurate result spans more than an order of magnitude, roughly from $3\times T_\tau $ up to $60\times T_\tau$, when demanding
a 10\% accuracy.
Leptogenesis calculations that aim for a precision that is better than order one should therefore include these effects, where necessary.

\section{Partial Equilibration of the $b$-Quark Yukawa Interactions
during Leptogenesis}
\label{sec:Example:Leptogenesis}

In the previous Section~\ref{sec:partial:tauequilibration}, we have treated a 
single lepton flavour during the equilibration
of $\tau$-Yukawa interactions only, which cannot be justified for realistic scenarios.
A proper treatment would require to take account of lepton flavour decoherence
along the lines of Refs.~\cite{Beneke:2010dz,Blanchet:2011xq}, which is a considerable
complication beyond the scope of the present paper. Therefore,
we now consider Leptogenesis  form the out-of-equilibrium decay of a singlet 
Majorana neutrino $N_1$
with mass of order $10^{13} {\rm GeV}$, which is much above $T_\tau$ and hence
in the unflavoured regime, such that we can effectively consider the
production and the washout of a lepton asymmetry within a single linear
combination of $e,\mu,\tau$, that we refer to as $\ell_\parallel$. The discussion
and results of this Section should therefore be applicable to realistic scenarios.
The regime of partial equilibration of the $b$-quark Yukawa interactions overlaps
with those of the weak and the strong sphalerons, such that we need to consider
these effects in conjunction.

However, it is of interest that an intermediate enhancement of the freeze-out asymmetry
may also result from non-leptonic spectator processes, {\it i.e.}
from $b$-quark Yukawa and
strong sphaleron interactions only. We therefore first perform the numerical analysis
for a vanishing rate of the weak sphalerons.

In order to account for the partial equilibration of
the $b$-quark Yukawa interactions and the strong sphaleron
processes, we need to identify linear combinations of charges that are only violated by
these particular interactions. We notice that $q_{Q1}$, the charge density
of a single component in the left-handed quark multiplet of the first generation
is only altered by strong sphalerons, while $\Delta_{\rm down}=q_{b}-q_{d1}$
is only altered by $b$-quark Yukawa interactions. Of course, we could have
chosen to use $q_{Q2}$ and $q_{d2}$ instead of the first generation charge densities
at this point, {\it i.e.} it is understood that $q_{Q2}=q_{Q1}$ and
$q_{d2}=q_{d1}$. Moreover, as long as the second-generation quark Yukawa-couplings
are out of equilibrium, the charge densities for the right-handed up- and
down-type quarks of the first two generations must agree, {\it i.e.}
$q_{d1,2}=q_{u1,2}$, as we assume in the following.
We can therefore remove any explicit
reference to $q_{Q2}$ from the equations by replacing it
with $q_{Q1}$ and of $q_{d2}$, $q_{u1,2}$ by replacing these with
$q_{d1}$. Of course, the corresponding substitutions can be made
for the chemical potentials as well.

We account for the lepton number violating processes through the charge density
$\Delta_\parallel=B/3-2 q_{\ell\parallel}$ (with the factor two being due to the
fact that $\ell_\parallel$ accounts for one doublet component only), which is
strictly conserved by weak sphalerons, and where $B$ is the baryon number density.
While these are
out-of-equilibrium at the temperatures under consideration (and we could as
well use ${\ell_\parallel}$ instead of $\Delta_\parallel$), the choice of $\Delta_\parallel$
is useful in view of extending the present analysis to lower temperatures.

Defining $Y_{\Delta\parallel}=\Delta_\parallel/s$,
we arrive at the 
following network of Boltzmann equations:
\begin{subequations}
\label{kineqs:blep}
\begin{align}
\frac{d Y_{N1}}{dz}=&-\bar{\cal C}_{N1}(Y_{N1}-Y_{N1}^{\rm eq})\,,\\
\label{kineqs:lepto}
\frac{d Y_{\Delta\parallel}}{dz}=&-\bar S(Y_{N1}-Y_{N1}^{\rm eq})+2 \bar W\left(Y_{\ell\parallel}+\frac12 Y_{H}\right)\,,\\
\label{kineqs:blep:hb}
\frac{d Y_{\Delta{\rm down}}}{dz}=&-\bar\Gamma_b\left[Y_{b}-Y_{Q3}+\frac12 Y_H\right]\,,\\
\label{kineqs:blep:ss}
\frac{d Y_{Q1}}{dz}=&-\frac{\Gamma_{\rm ss}}{2 T^3}\frac{T_{\rm com}}{T M_1}
\left[6 Y_{Q3}-3 Y_{t}-3 Y_b+24 Y_{Q1}\right]\,.
\end{align}
\end{subequations}
Here $\bar\Gamma_b = h_b^2 \gamma^{\rm fl \delta d} T_{\rm com}/(T M_1)$, and factors of $  T_{\rm com}/(T M_1)$ again arise from the change of variables compared to (\ref{eqn:flavour_t},{\ref{eqn:strongSph}). 
In addition, the evolution of the density of $N_1$ is given by Eq.~(\ref{eqn:eom1}).
In the strong washout regime and
for $z\gg1$, we may again use the approximate solution~(\ref{av:rates}),
with $Y_1$ being the Yukawa coupling between $N_1$ and $\ell_\parallel$.

Now, we need to relate the quantities
$Y_{\Delta\parallel}$, $Y_{\Delta{\rm down}}$ and $Y_{Q1}$
that appear on the left-hand side of the
Boltzmann equations to
$Y_{\ell\parallel}$, $Y_{Q3}$,
$Y_t$, $Y_b$, and $Y_H$ to $Y_{\Delta\parallel}$ on the right-hand side.
The necessary constraints can be obtained following Ref.~\cite{Barbieri:1999ma},
by using the equilibrium conditions
\begin{subequations}
\begin{align}
\mu_{Q3}-\mu_{t{\rm R}}+\mu_H&=0\quad\textnormal{(top-quark Yukawa interactions)}\,,\\
\mu_{Q3}+2\mu_{Q1}+2\mu_{t{\rm R}}+4\mu_{d1}&\notag\\
-\mu_{b{\rm R}}-2\mu_{d1}
-\mu_{\ell\parallel}+2
\mu_H&=0\quad \textnormal{(weak hypercharge neutrality)}\,,\\
\label{B:conservation}
2\mu_{Q3}+\mu_{t\rm R}+\mu_{b \rm R}+4\mu_{Q1}+4\mu_{d1}&=0\quad\textnormal{($B$ conservation)}\,, \\
\label{ss:eq}
2 \mu_{Q3} +\mu_{t\rm R} + \mu_{b\rm R} - 2\mu_{Q1} -2 \mu_{d1}& = 0 \quad\textnormal{(equal quark flavour asymmetries)}\,, 
\end{align}
\end{subequations}
where we have dropped chemical potentials that are identically zero. Note that
Eq.~(\ref{ss:eq}) follows from the flavour-diagonal nature of strong sphaleron
interactions. The last two equations immediately imply $\mu_{Q1} = -\mu_{d1}$, which has implicitly been used above to simplify the right-hand side of Eqn.~(\ref{kineqs:blep}d). 
For the remaining quantities, we find
\begin{align}
\label{AMatrix:Yb:sph}
\left(Y_{\ell\parallel},Y_{Q3},Y_t,Y_{b},Y_H\right)^t
=A(Y_{\Delta\parallel},Y_{\Delta{\rm down}},Y_{Q1})^t\,,
\end{align}
where
\begin{align}
A=
\left(
\begin{array}{ccc}
-\frac12 & 0 & 0\\
\frac{1}{18} & -\frac59 & \frac59\\
-\frac19 & \frac19 & -\frac19\\
0 & 1 & -1\\
-\frac13 & \frac43 & -\frac43
\end{array}
\right)
\,.
\end{align}
Note that this matrix acts on the space of charge densities defined by Eq.~(\ref{AMatrix:Yb:sph})
and is therefore a different quantity from the matrix $c_\ell$ and the vector $c_H$,
that occur for flavoured Leptogenesis and act on the space of SM lepton
flavours~\cite{Barbieri:1999ma,Nardi:2006fx,Abada:2006fw}.

As can be verified explicitly from the numerical results,
it turns out that the strong sphalerons have only a small effect on the charge
redistribution, what alleviates the systematic uncertainties from the strong
sphaleron rate to some extent. It is therefore a good approximation to take out
Eq.~(\ref{kineqs:blep:ss}) from the system of kinetic equations~(\ref{kineqs:blep})
and replace it by the equilibrium condition
\begin{align}
\label{ss:eq:constraint}
2\mu_{Q3}-\mu_{t{\rm R}} - \mu_{b \rm{R}} +4\mu_{Q1}-4\mu_{d1}&=0\quad\textnormal{(strong sphalerons)}\,.
\end{align}
In that case, the equilibrium conditions relate the charges that
appear in the kinetic equations as
\begin{align}
\left(Y_{\ell\parallel},Y_{Q3},Y_{b},Y_H\right)^t
=A(Y_{\Delta\parallel},Y_{\Delta{\rm down}})^t\,,
\end{align}
where
\begin{align}
A=
\left(
\begin{array}{ccc}
-\frac12 & 0\\
\frac{1}{23} & -\frac{10}{23}\\
\frac{1}{46} & \frac{18}{23}\\
-\frac{7}{23} & \frac{24}{23}
\end{array}
\right)
\,.
\end{align}

\begin{figure}[t]
\begin{center}
\includegraphics[width=0.45\textwidth]{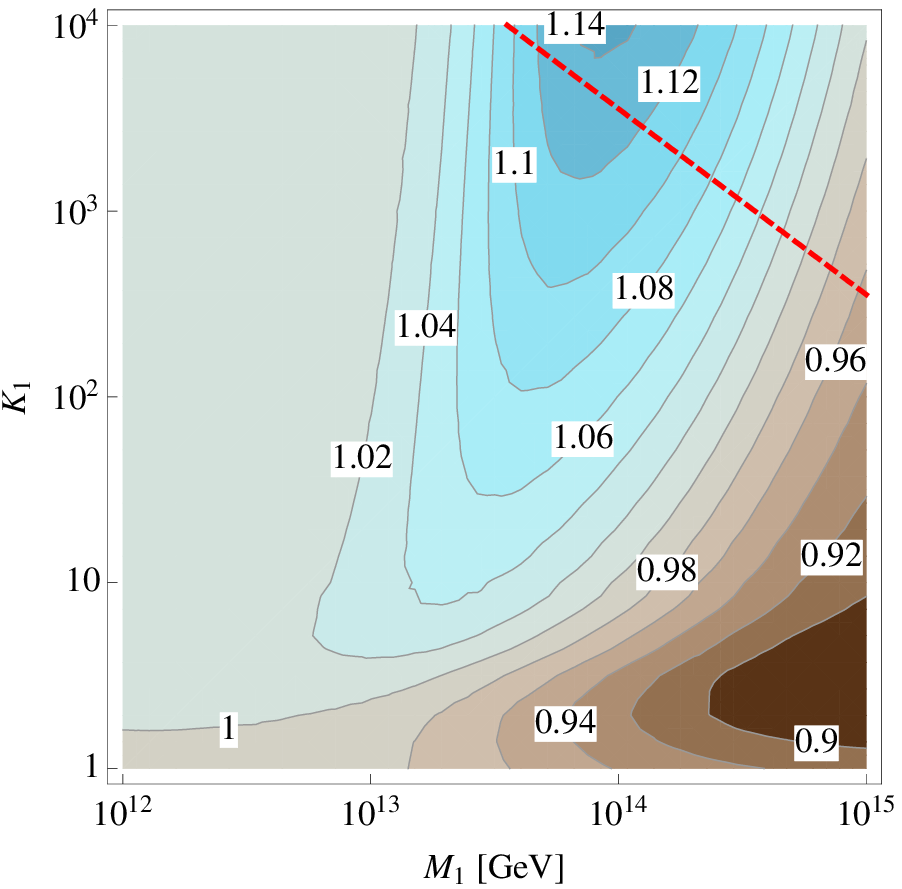}
\hspace*{.5cm}
\includegraphics[width=0.45\textwidth]{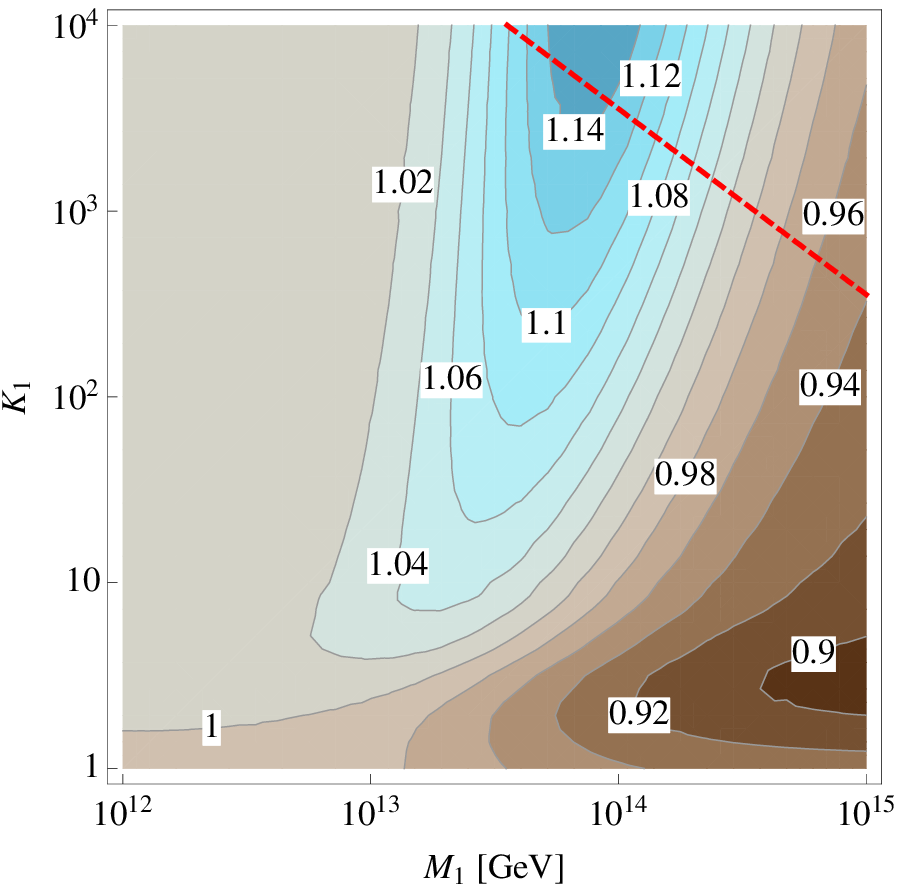}
\caption{\label{fig:strongsphbYukawa}Left: Enhancement of the asymmetry from partial equilibration
of $b$-quark Yukawa interactions and strong sphalerons
as function of the lightest right-handed neutrino mass $M_1$ and of the washout strength $K_1$ as defined in the text. Right: Same as left, but with the strong sphaleron imposed
to be in equilibrium. As before, in the regions above the dashed red lines $Y_1$ is non-perturbatively large. }
\end{center}
\end{figure}

Finally in order to quantify the importance of the partial equilibration, we also need to consider the case where the bottom Yukawa interactions and strong sphalerons
are fully equilibrated. As stated above, fast strong sphaleron
processes impose the constraint~(\ref{ss:eq:constraint}), which replaces the
differential equation~(\ref{kineqs:blep:ss}).
For fully equilibrated bottom Yukawa interactions, there is the additional the constraint 
\begin{align}
	\mu_{Q3} - \mu_{bR} - \mu_H = 0\,,
\end{align}
that can be used to replace the differential equation~(\ref{kineqs:blep:hb}). This
implies that all asymmetries can be expressed in terms of $Y_{\Delta\parallel}$
only, and we can solve Eq.~(\ref{kineqs:lepto}) using the relation
\begin{align}
	\left( Y_{L\parallel}, Y_H \right)^t = \left(-\frac{1}{2}, - \frac{1}{5} \right)^t Y_{\Delta \parallel} \,. 
\end{align}
Note that these factors are different from the coefficients used in
Ref.~\cite{Nardi:2005hs}, since there it is assumed that both the bottom and tau Yukawa interactions equilibrate at the same time. 

In Figure~\ref{fig:strongsphbYukawa}, we show the ratio of the asymmetry computed when accounting
for the partial equilibration of the $b$-quark Yukawa interactions and strong sphalerons over
the asymmetry that results when assuming a full equilibration. In order to illustrate
the relatively small effect of the strong sphalerons, we include also the same plot
obtained when taking the strong sphaleron interactions to be in equilibrium.

Overall, the effect of partial bottom Yukawa equilibration is smaller than in the case of partial $\tau$ equilibration, exceeding the 10\% level only in small corners of parameter space. The reason for this is that in the limit where the bottom Yukawa is in equilibrium, the constraint becomes $Y_{\Delta {\rm down}} = Y_{\Delta \parallel}/10$, such that only a small fraction of the asymmetry is ever transferred into the quark flavour asymmetry $\Delta_{\rm down}$. In comparison, the equilibrium condition induced by the $\tau$ Yukawa corresponds to $Y_\ell = Y_R$, such that a larger fraction can be transferred and hidden.

\

Now, we turn to the realistic situation where weak sphalerons are active. Since we have
convinced ourselves about the small effect from strong sphalerons, we assume for simplicity that
these are fully equilibrated and account for the partial equilibration of $b$-quark Yukawa 
interactions and of weak sphalerons.

\begin{figure}[t]
\begin{center}
\includegraphics[width=0.45\textwidth]{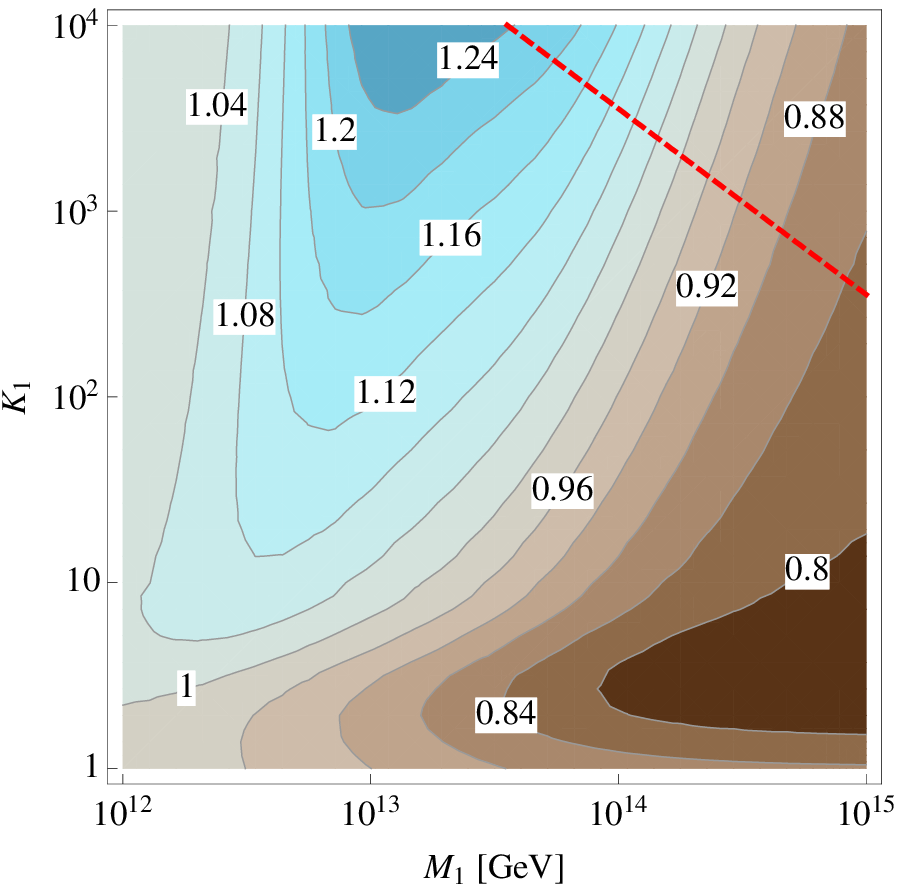}
\hspace*{.5cm}
\includegraphics[width=0.45\textwidth]{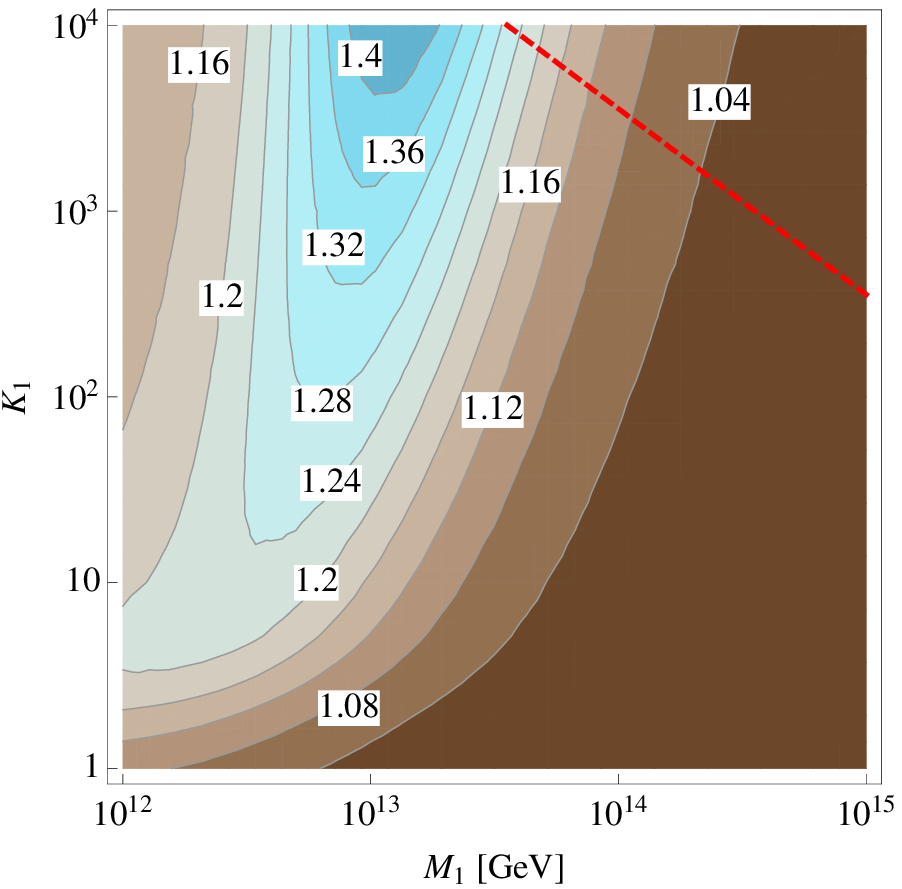}
\caption{\label{fig:weaksphbYukawa}Left: Ratio of the asymmetry when taking account of partial equilibration
of $b$-quark Yukawa interactions and weak sphalerons and the fully equilibrated approximation,
as function of the lightest right-handed neutrino mass $M_1$ and of the washout strength $K_1$ as defined in the text. Right: Ratio of the result obtained with partially equilibrated $b$-quarks and weak sphalerons divided by the result obtained with partially equilibrated $b$-quarks but neglecting weak sphaleron interactions.}
\end{center}
\end{figure}

Once the weak sphalerons become active, baryon number is not conserved anymore, such that the constraint~(\ref{B:conservation}) disappears and correspondingly a new variable appears in the Boltzmann equations. It is convenient to choose $q_{\ell_\perp}$, where
$\ell_\perp$ can be either of the two linear combination of SM lepton flavours that is
perpendicular to $\ell_\parallel$, as variable, since in the absence of charged lepton flavour effects this charge is only violated by weak sphalerons. The corresponding Boltzmann equation that describes the weak sphaleron is given by 
\begin{align}
\frac{dY_{\ell_\perp}}{dz}=-\frac{\Gamma_{\rm ws}}{2 T^3} \frac{T_{\rm com}}{T M_1} 
(9 Y_{Q3} + 18 Y_{Q1} + 3 Y_{\ell\parallel} + 6 Y_{\ell\perp})\,,
\end{align}
in addition to the equations for $Y_{\Delta \parallel}$ and $Y_{\Delta {\rm down}}$. 
With the strong sphalerons in equilibrium, the constraint~(\ref{ss:eq:constraint}) is imposed on the chemical potentials. Solving the new system of equations we can again express the fields appearing on the right-hand sides of the Boltzmann equations using the relations
\begin{align}
(Y_{\ell\parallel},Y_{\ell\perp},Y_{Q3},Y_b,Y_{Q1},Y_H)=A(Y_{\Delta\parallel},Y_{\ell_\perp},Y_{\Delta{\rm down}})^t\,,
\end{align}
where now
\begin{align}
A=
\left(
\begin{array}{ccc}
-\frac12 & 1 & 0\\
0 & 1 & 0\\
\frac{1}{23} & \frac12 & -\frac{10}{23}\\
\frac{1}{46} & \frac12 & \frac{18}{23}\\
-\frac{1}{46} & \frac12 & \frac{5}{23}\\
-\frac{7}{23} & 0 & \frac{24}{23}
\end{array}
\right)
\,,
\end{align}
and where $Y_{\ell_\perp} = q_{\ell_\perp}/s$ is the charge in a single gauge degree of freedom, as usual. 
The results for partially equilibrated $b$-quark Yukawa interactions and weak sphalerons are presented
in Figure~\ref{fig:weaksphbYukawa}. We observe that the partial equilibration
of weak sphaleron processes, that include leptons,
have a stronger enhancing effect than the non-leptonic processes considered above. Of course, this is
because the weak sphalerons are more efficient in hiding asymmetries due to their direct coupling to 
the initial asymmetry within $\ell_\parallel$.
When comparing the left panel of Figure~\ref{fig:weaksphbYukawa} with either of the plots in
Figure~\ref{fig:strongsphbYukawa} we observe that there is no clear separation of the regions where
the enhancement can be attributed to $b$-quark interactions from regions where it results from
weak sphalerons. Accounting for both interactions together apparently leads to a broadening of the region of enhancement. The purpose of the plot in the right panel of Figure~\ref{fig:weaksphbYukawa} is to isolate the effect from the weak sphalerons only, by dividing out the $b$-quark effects.
Finally, we mention that for vlaues at the lower end of the range of $M_1$ that is presented in
Figure~\ref{fig:weaksphbYukawa}, one should already expect effects from the $\tau$-lepton Yukawa 
coupling to be of importance. As this however also requires to include flavour decoherence, the
total effect will be more model-dependent, what should be subject of a future study.

\section{Discussion and Conclusions}
\label{sec:conclusions}

The following conclusions can be drawn
from the numerical examples presented in Sections~\ref{sec:partial:tauequilibration} and \ref{sec:Example:Leptogenesis}:
\begin{itemize}
\item
As one should expect, the equilibration temperatures ({\it cf.} Figure~\ref{fig:regions}
and Table~\ref{tab:rates}), defined by the time
when the respective reaction rates agree with the Hubble rate, indicate the
location of the parametric regime where the partial equilibration of spectator
effects is of importance. More precisely, in terms of the mass $M_1$
of the decaying right-handed neutrino, it ranges from roughly a factor
of $3$ above the equilibration temperature and spans two orders of magnitude,
when aiming to calculate the spectator effect with a (relative) accuracy at the 10\% level.
In terms of $M_1$, the partial equilibration regime is located above the equilibration
temperature, because the freeze-out occurs at temperatures below $M_1$, {\it cf.}
Eq.~(\ref{z:f}).
This is also a simple explanation of the shift of the partial equilibration regime toward
larger values of $M_1$ in the case of stronger washout, as $z_{\rm f}$ increases slowly
with $K_1$.
\item
For the two examples ($\tau$-Yukawa equilibration and $b$-Yukawa together with strong-sphaleron equilibration) worked out numerically in Sections~\ref{sec:partial:tauequilibration} and~\ref{sec:Example:Leptogenesis}, 
the freeze-out asymmetry assuming full equilibration
of the spectator fields is larger than when these effects are neglected, in consistency with the well-known results
presented in Refs.~\cite{Nardi:2005hs,Nardi:2006fx}.
For partial equilibration though, we find that the freeze-out asymmetry can be enhanced
compared to the two limiting cases, {\it cf.} Figures~\ref{fig:evolution}-\ref{fig:weaksphbYukawa}. This observation is best explained by the graphs in the
left panel of Figure~\ref{fig:evolution}.
The enhancement is typically more pronounced the
stronger the washout strength is. This is because during the early stages of Leptogenesis,
the large asymmetries that are present can be partly transferred to spectator fields.
When the spectators do not reach full chemical equilibrium with the fields $\ell$
or $H$ prior to freeze out, asymmetries can be partly hidden from washout.
Only after freeze out, chemical equilibration is reached such that these asymmetries
can be converted to a baryon asymmetry by weak sphaleron processes.
This ``hiding mechanism''  is different from and more efficient than the case of full equilibration, where the
asymmetries in the spectator fields track the evolution of the left-handed lepton asymmetries, what typically only results in a moderate reduction of the washout strength.
\end{itemize}

On possible future directions of research and applications of the present methods,
we make the following remarks:
\begin{itemize}
\item
In order to properly study the regime of partial equilibration of $\tau$- and  $\mu$-Yukawa
couplings, effects of partial flavour decoherence should be considered. A possible direction
would be to combine the methods presented here with the treatment of flavour effects
in Ref.~\cite{Beneke:2010dz}.
\item
An enhancement of the asymmetries compared to the limiting cases of complete equilibration
or complete absence of spectator effects is most pronounced for large washout strengths $K_1$.
Consequently, this requires large values of the Yukawa coupling $Y_1$ of the right-handed
neutrinos. From the standard parametrisation of these couplings in the type-I see-saw
mechanism~\cite{Casas:2001sr}, it can be seen that this is generally possible, but also
that this requires cancellations in order to suppress the masses of the SM neutrinos.
Scenarios with very strong washout can however be motivated and realised in the context of
resonant Leptogenesis by appealing to approximate
symmetries~\cite{Pilaftsis:2005rv}, such that the effects discussed in the present paper
may be particularly sizeable.
\item
Instead of the renormalisable or non-perturbative rates that scale $\propto T$,
it may be of interest to consider non-renormalisable interactions with rates
$\propto T^n$, where $n>1$, that mediate
the initial asymmetry to the spectator sector. Such processes are faster
at early times when large asymmetries are present,
such that these can be more efficiently hidden from strong washout.
This should allow for Baryo- and Leptogenesis from the out-of-equilibrium
decay of particles that are more strongly coupled than the right-handed neutrinos in
the standard scenarios of Leptogenesis, what may possibly lead to interesting phenomenological
prospects. We note in that context that the effect of protecting primordial asymmetries from
washout at later stages through what one may essentially call a non-renormalisable
spectator effect was first pointed out  in Ref.~\cite{Cline:1993bd}.
\item
The relations between the charge densities and chemical potentials receive
thermal radiative corrections, that would enter the present calculations as a
next-to leading-order effect, and that have been reported recently in Ref.~\cite{Bodeker:2014hqa}.
Due to the infrared population of bosonic fields, these corrections turn out
most sizeable for the Higgs boson. Future calculations with the goal of a high precision should also
include these effects.
\item
It will be of interest to investigate the dependence of the freeze-out asymmetry on the
initial conditions.
An important feature of Leptogenesis in the strong washout regime is that the freeze-out asymmetry is approximately
independent on the initial distribution of the sterile neutrinos. In the present work,
we always assume a thermal initial distribution for these particles. However, from
Figure~\ref{fig:evolution}, we can see that due to partially equilibrated spectator
particles, the outcome of Leptogenesis may be sensitive to the initial conditions
after all. A substantial part of the asymmetry is produced and partly transferred to
the spectators at small $z$.
When assuming non-thermal ({\it e.g.} vanishing) initial conditions for the
sterile neutrinos, the initial deviation from equilibrium $Y_{N1}-Y_{N1}^{\rm eq}$
will be enhanced compared to the thermal case, and consequently a larger asymmetry within
the spectator fields will be generated at these times.
We therefore
expect that in such a case the corrections to the standard calculations from partial equilibration of spectators will be more important than for the thermal initial conditions
considered in this paper.
\item
In connection with the previous point, it will be important to improve 
the predictions quantitatively by treating the relativistic regime $z\ll M_1$ more accurately.
For thermal initial distributions of the sterile neutrinos, we see from Figure~\ref{fig:evolution} that about half of the initial spectator asymmetry
is generated for $z\lsim0.5$, where the non-relativistic approximation for the
rate of sterile neutrino interactions become in principle inapplicable.
However, combining Eqs.~(\ref{S:average}) and~(\ref{DeltaYN}), we see that corrections to
the rate $\bar{\cal C}_{N1}(z)$ approximately cancel in the source term. Washout becomes quantitatively relevant
only for larger $z$, where the non-relativistic approximations used here are valid.
We therefore expect the present results not to change substantially upon improving on
the calculation of $\bar{\cal C}_{N1}(z)$ in the relativistic regime, $z\ll1$.
This may not be the case however when assuming vanishing initial conditions for
the sterile neutrinos, where the deviation from equilibrium for $z\ll1$ is large.
The pertinent $CP$ conserving rates are calculated in Refs.~\cite{Anisimov:2010gy,Besak:2012qm,Garbrecht:2013bia}, and it should be
straightforward to substitute these into the expressions for the $CP$-violating
source in Ref.~\cite{Beneke:2010wd} in the hierarchical limit.
When including spectator effects, it may therefore turn out that quantum statistical
corrections and relativistic rates are of importance for Leptogenesis in the strong
washout regime, in contrast to the most simple scenarios. We will investigate this
possibility in future work.
\end{itemize}

Whether spectator effects and their partial equilibration should be included in a particular
calculation of the lepton asymmetry is is a matter of the precision that it aims for. Since the prospects for a precise measurement of the high scale parameters relevant for Leptogenesis are rather bad, one may perhaps decide to neglect this effect in a pragmatic phenomenological study.
However once additional constraints on the parameter space are imposed, {\it e.g.} when one demands the washout of primordial asymmetries as in Ref.~\cite{DiBari:2014eqa} or imposes the gravitino bound on the reheating temperature~\cite{Khlopov:1984pf,Heisig:2013sva}, including these effects might be warranted or even necessary to gain a proper understanding of the viable parameter space. 
In addition, the insights gained may be helpful in
identifying alternative mechanisms of Baryogenesis, where of particular interest are scenarios
that are accessible to observational tests.

\subsection*{Acknowledgements}
We would like to thank M.~Garny for discussions and comments on the manuscript. 
BG acknowledges
support by the Gottfried Wilhelm Leibniz programme
of the DFG and by the DFG cluster of excellence `Origin and Structure of the Universe'.
\appendix

\section{Rate of Quark-Yukawa Mediated Processes}
\label{appendix:b:rate}

By appropriate replacements of coupling constants and group-theoretical factors,
the intermediate numerical results that are presented in Ref.~\cite{Garbrecht:2013bia}
can be used in order to derive the coefficient $\gamma^{{\rm fl}\delta d}$
for the equilibration rate of down-type Yukawa mediated processes.
It can be decomposed as
\begin{align}
\label{gamma:d}
\gamma^{{\rm fl}\delta d}=\gamma^{{\rm fl}(\phi)\delta d}
+\gamma^{{\rm fl}(\phi)\delta d}
+\gamma^{{\rm fl}(Q)\delta d}
+\gamma^{{\rm fl}(d)\delta d}
+\gamma^{{\rm fl}\delta d}_{\rm vertex}
+\gamma^{{\rm fl}\delta d}_{1\leftrightarrow 2}
\,,
\end{align}
where the individual terms are given by
\begin{subequations}
\label{gamma:d:individuals}
\begin{align}
\label{gamma:phi:d}
\gamma^{{\rm fl}(\phi)\delta d}=&
7.71\times10^{-4}G^{(\phi)}T
+1.32\times10^{-3}h_t^2 T\,,\\
\label{gamma:Qd:d}
\gamma^{{\rm fl}(Q,d)\delta d}=&
3.72\times10^{-3}G^{(Q,d)} T-8.31\times10^{-4}G^{(Q,d)}T\log G^{(Q,d)}\,,\\
\label{gamma:vertex:d}
\gamma^{{\rm fl}\delta d}_{{\rm vertex}}=&
-7.72\times10^{-4}\left(\frac{8}{3} g_3^2 +\frac32 g_2^2 +\frac{5}{18} g_1^2\right)T\,,\\
\label{gamma:LPM:d}
\gamma^{{\rm fl}\delta d}_{1\leftrightarrow 2}=&
1.7\times 10^{-3}\left(\frac83 g_3^2+\frac 32 g_2^2\right)T\,,
\end{align}
\end{subequations}
with
\begin{subequations}
\begin{align}
G^{(\phi)}=&\frac32 g_2^2+\frac 12 g_1^2\,,\\
G^{(Q)}=&\frac83 g_3^2+\frac32 g_2^2+\frac{1}{18}g_1^2\,,\\
G^{(d)}=&\frac83 g_3^2 +\frac29 g_1^2\,.
\end{align}
\end{subequations}
While the results~(\ref{gamma:phi:d},\ref{gamma:Qd:d},\ref{gamma:vertex:d}) are accurate to leading order in the couplings
following the calculation of Ref.~\cite{Garbrecht:2013bia}, the
contribution~(\ref{gamma:LPM:d}) corresponds to an estimate.
A full calculation along the lines of Refs.~\cite{Anisimov:2010gy,Arnold:2002ja}
may turn out to be somewhat involved, but we expect that Eqs.~(\ref{gamma:d:individuals}) should
be suitable for the present purposes, in particular because the dominating
exchange of quarks in the $t$-channel is accurately captured through
Eq.~(\ref{gamma:Qd:d}) to leading order. With the
values for $g_2$ and $g_3$ that are stated in Section~\ref{section:regimes}
and $g_1=0.41$ for temperatures of order $10^{12}\,{\rm GeV}$,
Eq.~(\ref{gamma:d}) yields $\gamma^{{\rm fl} \delta d} \approx0.01$.

For completeness, we also estimate the rate for up-type Yukawa mediated interactions. Here the term proportional to $h_t^2$ in~(\ref{gamma:phi:d}) does not contribute, since the Yukawa coupling alone does not break left-handed quark number (it only mediates processes like $Q \bar{Q} \leftrightarrow t_R \bar{t}_R$ or $Q \bar{t}_R \leftrightarrow Q \bar{t}_R$, but not e.g. $Q\bar{t}_R \leftrightarrow \bar{Q} t_R$). The vertex contribution becomes 
\begin{align}
	\gamma^{{\rm fl}\delta u}_{{\rm vertex}}=&
-7.72\times10^{-4}\left(\frac{8}{3} g_3^2 +\frac32 g_2^2 +\frac{17}{18} g_1^2\right)T\,,
\end{align}
and $G^{(d)}$ should be replaced with 
\begin{align}
	G^{(u)} & = \frac 83 g_3^2 + \frac{8}{9} g_1^2 \,.
\end{align}
Altogether we obtain $\gamma^{{\rm fl} \delta u} \approx 0.01~T$. The rates for up and down type quarks differ in the next digit, but this is beyond the level of accuracy of this estimate. For the charged lepton Yukawa interaction rates, the running of $g_2$ and $g_1$ induces a temperature dependence which is negligible for all practical purposes. On the other hand the QCD coupling runs more strongly, and the rates $\gamma^{{\rm fl}\delta u} \approx \gamma^{{\rm fl}\delta d}$ increase to $0.012~T$ at $T \approx 10^9$\;GeV and to $0.015~T$ at $T\approx 10^6$\;GeV. This effect is included in our estimates of the quark Yukawa equilibration temperatures in Tab.~\ref{tab:rates}.


\begin{thebibliography}{99}
\bibliographystyle{unsrt}

\bibitem{Fukugita:1986hr}
  M.~Fukugita and T.~Yanagida,
  ``Baryogenesis Without Grand Unification,''
  Phys.\ Lett.\ B {\bf 174} (1986) 45.

\bibitem{Nardi:2005hs}
  E.~Nardi, Y.~Nir, J.~Racker and E.~Roulet,
  ``On Higgs and sphaleron effects during the leptogenesis era,''
  JHEP {\bf 0601} (2006) 068
  [hep-ph/0512052].

  
\bibitem{Nardi:2006fx}
  E.~Nardi, Y.~Nir, E.~Roulet and J.~Racker,
  ``The importance of flavor in leptogenesis,''
  JHEP {\bf 0601}, 164 (2006)
  [arXiv:hep-ph/0601084].
  
\bibitem{Buchmuller:2001sr}
  W.~Buchm\"uller and M.~Pl\"umacher,
  ``Spectator processes and baryogenesis,''
  Phys.\ Lett.\ B {\bf 511} (2001) 74
  [hep-ph/0104189].

\bibitem{Barbieri:1999ma}
  R.~Barbieri, P.~Creminelli, A.~Strumia and N.~Tetradis,
  ``Baryogenesis through leptogenesis,''
  Nucl.\ Phys.\ B {\bf 575} (2000) 61
  [hep-ph/9911315].

\bibitem{Abada:2006fw}
  A.~Abada, S.~Davidson, F.~X.~Josse-Michaux, M.~Losada and A.~Riotto,
  ``Flavour Issues in Leptogenesis,''
  JCAP {\bf 0604}, 004 (2006)
  [arXiv:hep-ph/0601083].

  
\bibitem{Beneke:2010dz}
  M.~Beneke, B.~Garbrecht, C.~Fidler, M.~Herranen and P.~Schwaller,
  ``Flavoured Leptogenesis in the CTP Formalism,''
  Nucl.\ Phys.\  B {\bf 843} (2011) 177
  [arXiv:1007.4783 [hep-ph]].
  
\bibitem{Anisimov:2010gy}
  A.~Anisimov, D.~Besak and D.~B\"odeker,
  ``Thermal production of relativistic Majorana neutrinos: Strong enhancement by multiple soft scattering,''
  JCAP {\bf 1103} (2011) 042
  [arXiv:1012.3784 [hep-ph]].

\bibitem{Besak:2012qm}
  D.~Besak and D.~B\"odeker,
  ``Thermal production of ultrarelativistic right-handed neutrinos: Complete leading-order results,''
  JCAP {\bf 1203} (2012) 029
  [arXiv:1202.1288 [hep-ph]].

\bibitem{Garbrecht:2013bia}
  B.~Garbrecht, F.~Glowna and P.~Schwaller,
  ``Scattering Rates For Leptogenesis: Damping of Lepton Flavour Coherence and Production of Singlet Neutrinos,''
  Nucl.\ Phys.\ B {\bf 877} (2013) 1
  [arXiv:1303.5498 [hep-ph]].
  
\bibitem{Bodeker:1998hm}
  D.~B\"odeker,
  ``On the effective dynamics of soft nonAbelian gauge fields at finite temperature,''
  Phys.\ Lett.\ B {\bf 426} (1998) 351
  [hep-ph/9801430].

\bibitem{Moore:2000mx}
  G.~D.~Moore,
  ``Sphaleron rate in the symmetric electroweak phase,''
  Phys.\ Rev.\ D {\bf 62} (2000) 085011
  [hep-ph/0001216].
  
\bibitem{Garny:2009rv}
  M.~Garny, A.~Hohenegger, A.~Kartavtsev and M.~Lindner,
  ``Systematic approach to leptogenesis in nonequilibrium QFT: vertex
  contribution to the CP-violating parameter,''
  Phys.\ Rev.\  D {\bf 80} (2009) 125027
  [arXiv:0909.1559 [hep-ph]].
 
\bibitem{Beneke:2010wd}
  M.~Beneke, B.~Garbrecht, M.~Herranen and P.~Schwaller,
  ``Finite Number Density Corrections to Leptogenesis,''
  Nucl.\ Phys.\  B {\bf 838} (2010) 1
  [arXiv:1002.1326 [hep-ph]].
  
\bibitem{Anisimov:2010dk}
  A.~Anisimov, W.~Buchmuller, M.~Drewes and S.~Mendizabal,
  ``Quantum Leptogenesis I,''
  Annals Phys.\  {\bf 326} (2011) 1998
  [arXiv:1012.5821 [hep-ph]].
  
\bibitem{Buchmuller:2004nz}
  W.~Buchm\"uller, P.~Di Bari and M.~Pl\"umacher,
  ``Leptogenesis for pedestrians,''
  Annals Phys.\  {\bf 315} (2005) 305
  [hep-ph/0401240].
  
\bibitem{Blanchet:2011xq}
  S.~Blanchet, P.~Di Bari, D.~A.~Jones and L.~Marzola,
  ``Leptogenesis with heavy neutrino flavours: from density matrix to Boltzmann equations,''
  JCAP {\bf 1301} (2013) 041
  [arXiv:1112.4528 [hep-ph]].
  
  
\bibitem{Moore:2000ara}
  G.~D.~Moore,
  ``Do we understand the sphaleron rate?,''
  hep-ph/0009161.

\bibitem{D'Onofrio:2012jk}
  M.~D'Onofrio, K.~Rummukainen and A.~Tranberg,
  ``The Sphaleron Rate through the Electroweak Cross-over,''
  JHEP {\bf 1208} (2012) 123
  [arXiv:1207.0685 [hep-ph]].

\bibitem{Khlebnikov:1988sr}
  S.~Y.~Khlebnikov and M.~E.~Shaposhnikov,
  ``The Statistical Theory of Anomalous Fermion Number Nonconservation,''
  Nucl.\ Phys.\ B {\bf 308} (1988) 885.


\bibitem{Burnier:2005hp}
  Y.~Burnier, M.~Laine and M.~Shaposhnikov,
  ``Baryon and lepton number violation rates across the electroweak crossover,''
  JCAP {\bf 0602} (2006) 007
  [hep-ph/0511246].

\bibitem{Bento:2003jv}
  L.~Bento,
  ``Sphaleron relaxation temperatures,''
  JCAP {\bf 0311} (2003) 002
  [hep-ph/0304263].
  
\bibitem{Moore:1997im}
  G.~D.~Moore,
  ``Computing the strong sphaleron rate,''
  Phys.\ Lett.\ B {\bf 412} (1997) 359
  [hep-ph/9705248].
  
\bibitem{Casas:2001sr}
  J.~A.~Casas and A.~Ibarra,
  ``Oscillating neutrinos and muon to e, gamma,''
  Nucl.\ Phys.\ B {\bf 618} (2001) 171
  [hep-ph/0103065].
  
\bibitem{Pilaftsis:2005rv}
  A.~Pilaftsis and T.~E.~J.~Underwood,
  ``Electroweak-scale resonant leptogenesis,''
  Phys.\ Rev.\ D {\bf 72} (2005) 113001
  [hep-ph/0506107].

\bibitem{Cline:1993bd}
  J.~M.~Cline, K.~Kainulainen and K.~A.~Olive,
  ``Protecting the primordial baryon asymmetry from erasure by sphalerons,''
  Phys.\ Rev.\ D {\bf 49} (1994) 6394
  [hep-ph/9401208].
  
 
\bibitem{Bodeker:2014hqa}
  D.~B\"odeker and M.~Laine,
  ``Kubo relations and radiative corrections for lepton number washout,''
  arXiv:1403.2755 [hep-ph].
  
\bibitem{DiBari:2014eqa}
  P.~Di Bari, S.~King and M.~Re Fiorentin,
  ``Strong thermal leptogenesis and the absolute neutrino mass scale,''
  arXiv:1401.6185 [hep-ph].

\bibitem{Khlopov:1984pf}
  M.~Y.~.Khlopov and A.~D.~Linde,
  ``Is It Easy to Save the Gravitino?,''
  Phys.\ Lett.\ B {\bf 138} (1984) 265.

\bibitem{Heisig:2013sva}
  J.~Heisig,
  ``Gravitino LSP and Leptogenesis after the first LHC results,''
  arXiv:1310.6352 [hep-ph].

\bibitem{Arnold:2002ja}
  P.~B.~Arnold, G.~D.~Moore and L.~G.~Yaffe,
  ``Photon and gluon emission in relativistic plasmas,''
  JHEP {\bf 0206} (2002) 030
  [hep-ph/0204343].

\end{thebibliography}
\end{document}